\documentstyle[12pt]{article}
\title{Bose-Einstein correlations}
\author{Kacper Zalewski\thanks{Supported in part by the KBN grant
2P03B 086 14}\\
Jagellonian University\\ and\\ Institute of Nuclear Physics, Krak\'ow, Poland}

\begin{document}
\maketitle
\begin{abstract}

The effect of Bose-Einstein correlations on multiplicity distributions of
identical pions is discussed. It is found that these correlations affect
significantly the observed multiplcity distributions, but Einstein's
condensation is unlikely to be achieved, unless "cold spots", i.e. regions,
where groups of pions with very small relative momenta are produced, occur in
high energy heavy-ion collisions.

\end{abstract}

\section{INTRODUCTION}

Bose-Einstein correlations are important in a variety of physical problems. For
a recent review cf. \cite{BAY}. In the present report I will review some work
on such correlations done recently in collaboration with A. Bia\l{}as. Let us
start with some general remarks.

In particle physics Bose-Einstein correlations imply for identical bosons an
apparent attraction in momentum space. Since the identical bosons (say $\pi^+$
mesons) are usually only a subset of the particles present in the final state
of a high energy collision, it is in general not possible to describe them by a
wave function. Instead one must use a density matrix $\rho (q,q\prime)$, where
$ q = (\vec{q}_1,\ldots,\vec{q}_N)$ denotes the set of momenta of all the
identical bosons and similarly for $q\prime$, or a generalized Wigner function
$W(P,X)$ cf. e.g. \cite{PR1}, \cite{PGG}, \cite{BKR} related to the density
matrix in the momentum representation by the formula

\begin{equation}
\rho_N(P+\frac{Q}{2},P-\frac{Q}{2}) = \int d^{4N}X\;W(P,X)e^{-iQX}.
\end{equation}

If $\rho^{(0)}_N$ is the density matrix for $N$ very similar but
distinguishable particles, the density matrix for the same particles with the
additional assumption that they are indistinguishable is (cf.
\cite{BZ1},\cite{BZ2},\cite{KZ1} and references given there)

\begin{equation}
\rho_N^{(0)}(p,q) =\sum_\sigma
\rho_N^{(0)}(p, q_\sigma),
\end{equation}
where $\sigma$ is a permutation of the momenta $\vec{q}_1,\ldots,\vec{q}_N$ and
the summation extends over all the $N!$ permutations $\sigma$. Assuming as
usual that $Tr \rho_N^{(0)} = 1$, one finds $Tr \rho_N \neq 1$. Thus, the
probability of producing $N$ identical bosons changes from $P^0(N)$ to ${\cal
N}P^0(N)Tr\rho_N$, where ${\cal N}$ is a normalization constant. This can lead
to spectacular changes in the multiplicity distribution \cite{PR2}.

We limit our discussion to independent particle production models, where the
multiplicity distribution before taking into account the identity of particles
is Poissonian, characterized by the parameter $\nu$

\begin{equation}
P^{(0)}(N) = \frac{\nu^N}{N!} e^{-\nu}
\end{equation}
and for a given particle multiplicity $N$ the density matrix (for
distinguishable particles) is a product of identical single particle density
matrices

\begin{equation}
\rho^{(0)}_N(p,q) = \prod_{i=1}^N \rho_1^{(0)}(p_i,q_i).
\end{equation}
We make no specific assumptions about the single particle density matrix,
which is assumed in its most general form

\begin{equation}
\rho_1^{(0)}(p,q) = \sum_n \langle p|n\rangle \lambda_n \langle n|q \rangle .
\end{equation}

Special cases of this model have been considered by many authors. A Gaussian
$\rho_1^{(0)}$ has been particularly popular, because for it the
symmetrization procedure can be performed explicitly.  The original approach
\cite{PR2},\cite{CZI} is, however, quite complicated. We will show that it can
be replaced by a very simple one, when the problem is translated into the
language of thermodynamics.

\section{ TRANSLATION INTO THERMODYNAMICS I: DISTINGUISHABLE PARTICLES}

As a warming up exercise and also in order to introduce the notation let us
consider first the problem of distinguishable particles.  The main idea
\cite{BZ3} is to interpret the single particle density operator as the
canonical density operator known from statistical physics

\begin{equation}
\hat{\rho}^{(0)}_1 = \sum_n |n\rangle \lambda_n \langle n| = \frac{1}{Z}
e^{-\beta \hat{H}},
\end{equation}
where the Hamiltonian $\hat{H} = \sum_n |n\rangle \varepsilon_n \langle n|$
and $Z$ is a normalizing constant. By comparison

\begin{equation}
\lambda_n = \frac{1}{Z} e^{-\beta \varepsilon_n}
\end{equation}
and from the condition $Tr \rho_1^{(0)} = 1$ one finds that $Z$ is the standard
partition function $\sum_n e^{-\beta \varepsilon_n}$. Thus $\lambda_n$ is just
the canonical probability that the particle is in state $n$.

The Hamiltonian corresponding to a given density matrix may be quite awkward,
but simple cases also occur. E.g. for a Gaussian density matrix

\begin{equation}
\rho_1^{(0)}(p,q) = \frac{1}{\sqrt{2\pi \Delta^2}} \exp\left[
-\frac{q_+^2}{2\Delta^2} - \frac{R^2}{2}q_-^2\right],
\end{equation}
where $q_+ = \frac{1}{2}(p+q)$ and $q_- = p-q$, the corresponding Hamiltonian
is that of a harmonic oscillator with\footnote{The formula below corresponds to
formula (13) of \cite{BZ3}, which has been misprinted there and should be
corrected as give below.}

\begin{equation}
m\omega = \frac{\Delta}{R}
\end{equation}
and

\begin{equation}
\hbar\omega = -\frac{1}{\beta}\log\frac{2R\Delta-1}{2R\Delta+1}.
\end{equation}
There is no problem at $R\Delta \leq \frac{1}{2}$, because from the Heisenberg
uncertainty principle $R\Delta \geq \frac{1}{2}$ and one can check that for
$R\Delta \rightarrow \frac{1}{2}$ also $\beta \rightarrow \infty$ and
$\hbar\omega$ has a finite limit. Note that according to (10) the frequency of
the oscillator is temperature dependent. Since, however, we are using
thermodynamics only as a formalism, this does not cause any trouble. For the
same reason our use of the canonical distribution does not mean that the system
of pions is in equilibrium.

\section{TRANSLATION INTO THERMODYNAMICS II: INDISTINGUISHABLE PARTICLES}

For indistinguishable particles the choice of a single particle as a
well-defined subsystem is impossible.  The procedure recommended by
thermodynamics textbooks is to choose instead all the particles occupying a
single particle state. Thus, for each single particle state $k$ there is a
subsystem with its state defined by the population of state $k$. The states of
the subsystem can be labelled by $n_k = 0,1,\ldots$. The corresponding energies
are, of course, $n_k\varepsilon_k$. The subsystem is open, i.e. it can exchange
particles with other subsystems. Consequently, the probability of state $n_k$ of
subsystem $k$ is given by the grand-canonical formula

\begin{equation}
P_k(n_k) = \frac{1}{{\cal Z}_k} e^{\beta(\mu - \varepsilon_k)n_k},
\end{equation}
where $\mu$ is known as the chemical potential and ${\cal Z}_k$ - the grand
partition function - is a normalization constant fixed by the condition
$\sum_{n_k=0}^\infty P_k(n_k) = 1$. By summing the corresponding geometrical
progression one finds

\begin{equation}
{\cal Z}_k = \frac{1}{1 - e^{\beta(\mu - \varepsilon_k)}}.
\end{equation}
The quantity $e^{\beta\mu}$ is known in thermodynamics as fugacity and in our
model
it equals $\frac{\nu}{Z}$. The probability distribution for the whole system
can be obtained by multiplying the probability distributions for the subsystems

\begin{equation}
P(n_0,n_1,\ldots) = P_0(n_0)P_1(n_1)\ldots .
\end{equation}
These formulae yield all that is necessary. We present two examples.

The average number of particles in subsystem $k$ is given by the textbook
formula

\begin{equation}
\langle n_k \rangle = \frac{1}{\beta} \frac{\partial \log{\cal
Z}_k}{\partial\mu}
= \frac{\nu \lambda_k}{1 - \nu\lambda_k}.
\end{equation}
For $\nu\lambda_k$ small the deviation of the denominator from unity can be
neglected, and the canonical formula valid for distinguishable particles is
reproduced. When, however, $\nu\lambda_0$, where $\lambda_0$ denotes the
largest eigenvalue of the density matrix, tends to one, $n_0 \rightarrow
\infty$, while the average population of the states corresponding to $\lambda_k
< \lambda_0$ tend to finite limits. This phenomenon is known as Einstein's
condensation. It was noticed in the Gaussian model by Pratt, who called such a
system a pion laser \cite{PR2}.

Let us calculate now the probability that among the $N$ pions produced  in an
interaction there are no $\pi^0$-s. The result is of interest, because cosmic
ray physicists report observations of high multiplicity events without
$\pi^0$-s \cite{LFH}. Such events are known as centauros. The probability of no
$\pi^0$-s in subsystem $k$ is

\begin{equation}
P_k(0) = \frac{1}{\cal Z}_k = \frac{1}{\langle n_k\rangle + 1}.
\end{equation}
Let us notice two limiting cases. For $\langle n_k \rangle \ll 1$ the formula
can be exponentialized and $P_k(0) \approx e^{-\langle n_k\rangle}$. For
$\langle n_k \rangle \gg 1$, $P_k(0) \approx \frac{1}{\langle n_k \rangle}$. No
$\pi^0$-s in the whole system means no $\pi^0$-s in every one of the subsystems
$k$. Multiplying the corresponding probabilities one finds, when all $\langle
n_k \rangle \ll 1$,

\begin{equation}
P(0) \approx e^{-N},
\end{equation}
where $N = \sum_k \langle n_k\rangle$. In the case when almost all the
particles are in the $k=0$ state

\begin{equation}
P(0) \approx \frac{1}{N}.
\end{equation}
The probability (16) is negligible for $N$ of the order of hundred or more,
but probability (17) is then much larger. It could explain the centauro
events, if the $\pi^0$-s were condensed. Thus Einstein's condensation could
lead to interesting phenomena (there are more cf. e.g. \cite{KZ1}). The problem
is, however, are we likely to reach condensation in realistic experimental
situations.

\section{COLD SPOTS}

Let us start with a simpler question: are the effects of symmetrization on the
multiplicity distribution large enough to be observable. This question must be
made quantitative. Suppose the effects are observable, when the population of
the $k=0$ state, corresponding to the largest eigenvalue $\lambda_0$, is
doubled due to symmetrization, i.e. when $\nu\lambda_0 \geq \frac{1}{2}$. For
the Gaussian model this condition can be rewritten as

\begin{equation}
\nu \geq \frac{1}{2}(R\Delta + \frac{1}{2})^3.
\end{equation}
For $R\Delta = 1$ this corresponds to $\nu \geq 1.7$. Using formula (14) summed
over $k$ one finds for $\pi^+$ mesons the equivalent condition $N_{\pi^+}
\geq 3.4$, which for the total number of pions yields $N_\pi \approx 3
N_{\pi_+} \geq 10$. This is the number of pions, which should be produced in a
so-called coherence region, i.e. within a region of space-time, where the
momenta of the pions produced are sufficiently close to each other to make
interference effective. As illustrated e.g. by formula (8) the contribution
from pion pairs, where $|q^2_-|$ is large, is negligible. This is in fact the
reason, why the production regions deduced from studies of Bose-Einstein
correlations are approximately spherical, while the full production region is
believed to be a string, or a bunch of strings. The condition $N_\pi \geq 10$
is not very restrictive and, therefore, we do expect the multiplicity
distribution of pions to be significantly modified by symmetrization. This has
been confirmed by much more realistic calculations \cite{JZA}, \cite{FWI},
where the unsymmetrized distributions had been calculated from LUND models and
then additional weights corresponding to symmetrization were introduced. This
observation is interesting, because it shows that the success of the LUND model
without symmetrization in reproducing multiplicity distributions of identical
pions is due to a cancellation of errors. First the parameters of the
unsymmetrized model are adjusted so that it reproduces correctly the observed
multiplicity distribution, which it should not, and then the significant
correction resulting from symmetrization is not included, so that the final
result is satisfactory.

Let us consider now the question, whether condensation of pions is expected to
occur in real life. As the quantitative condition we require that the $k=0$
state contains ten times more particles that the next (triple degenerate
according to the Gaussian model) state.  In the Gaussian model this
approximately corresponds to

\begin{equation}
N_{\pi^+} > 30 (R\Delta - \frac{1}{2}).
\end{equation}
Let us try again $R\Delta = 1$, which corresponds to $N_{\pi^+} > 15$. A
typical transverse momentum component for pions is $0.25$ GeV. Since $\Delta$
is equal to the root-mean-square momentum component, we put $\Delta = 0.25$ GeV
and consequently $R = 4$ GeV$^{-1} \approx 0.8$ fm. Thus, in the volume of,
say, $\frac{4}{3}\pi R^3 \approx 2$ fm${}^3$ there should be on the average
more than 15 $\pi^+$-s, or more than 45 pions. Since the radius of a pion is of
the order of one fm, this seems a very large density. Alternatively we can
estimate the energy density. Each pion carries an energy of about
$\sqrt{m^2_\pi + 3\Delta^2} \approx 0.5$ GeV. The energy density corresponding
to the formation of the quark-gluon plasma is about $1$GeV/fm$^3$. Thus again
we find that the density is too high to be a reasonable starting point for a
free pion evolution.

Let us put, however, $\Delta = 0.1$ GeV. Then $R=2.0$ fm and there are $15$
$\pi^+$-s per $34$ fm$^3$, which is much more reasonable. The corresponding
energy per pion is about $0.2$ GeV. Thus, the energy density is about $0.1$
GeV/fm$^3$, safely below the density of the plasma phase. If this density is
still too high, one can go on reducing $\Delta$ and increasing $R$. We conclude
that, if $\Delta$ is as for typical pions, the condensation is unlikely. But,
if in a heavy ion collision a sufficiently large "cold spot" is formed with low
$\Delta$, i.e. with low temperature, or equivalently with low relative momenta
of the particles produced there, there may be condensation of the pions
produced in this spot and some interesting phenomena may become observable
\cite{BZ4}.

\end{document}